\documentclass[12pt]{article}
\usepackage[T2A]{fontenc}
\usepackage[english]{babel}

\usepackage{amsmath,amssymb}
\usepackage[mathscr]{eucal}

\newcommand{\pl}{\partial}
\newcommand{\ol}{\overline}
\newcommand{\plr}{\partial r}

\newcommand{\Dtph}{\Delta_{\Omega}}
\newcommand{\DD}{\mathcal{D}}
\newcommand{\HH}{\mathscr{H}}
\newcommand{\Hh}{H}
\newcommand{\Hcl}{\mathcal{H}}
\newcommand{\WW}{\mathscr{W}}

\newcommand{\RR}{\mathbb{R}}

\newcommand{\Sph}{\mathbb{S}}

\newcommand{\ve}{\varepsilon}
\newcommand{\YY}{\mathrm{Y}}

\begin{document}
\begin{center}
{\Large
Quantum Hamiltonian Eigenstates\\[2mm]
for a Free Transverse Field
}\\
\vspace{0.3cm}
T.~A.~Bolokhov\\
\vspace{0.3cm}
{\it St.\,Petersburg Department of V.\,A.\,Steklov Mathematical Institute\\
         Russian Academy of Sciences\\
         27 Fontanka, St.\,Petersburg, Russia 191023}
\end{center}

\begin{abstract}
We demonstrate that the quantum Hamiltonian operator for a free transverse field
within the framework of the second quantization reveals an alternative set of
states satisfying the eigenstate functional equations.
The construction is based upon extensions of the quadratic form of the
transverse Laplace operator which are used as a source of spherical basis
functions with singularity at the origin.
This basis then naturally takes place of the one of plane or spherical waves
in the process of Fourier or spherical variable separation.
\end{abstract}

\section*{Introduction}
	The second quantization approach
\cite{Dirac}, \cite{Becchi}
	has been the framework for constructing the quantum field theory
	since the time of its inception in the first half of the 20th century.
	Later in the development, for the purpose of practical calculations
	of the scattering matrix elements, a wide recognition was given
	to the technique of Feynman's diagrams,
        which is based on the Lagrangian formulation of the classical theory.
	Unlike this latter technique, one of the advantages of the second
	(canonical) quantization is that it provides a description in terms
    of the quantum Hamiltonian operator.
	In a correctly defined quantum system the latter object must be
	a self-adjoint operator in Hilbert space.
	The finite-dimensional examples 
\cite{BF},
\cite{Jackiw}
	show that upon renormalization and the removal of singularities
	the Hamiltonian nominee may well be a symmetric but still not
	a self-adjoint operator representing
    a free particle on a restricted space of states.
	Such a candidate can be extended to a self-adjoint operator,
    however this procedure is ambiguous as it requires introduction
    of an extension parameter (the dimensional transmutation phenomenon as of
\cite{Jackiw},
\cite{LFres}).
    A similar effect is seemingly observed in the systems of infinite number
    of harmonic oscillators.
	We shall argue that the quadratic part of the quantum Hamiltonian of a free transverse vector field
\begin{equation*}
    \HH_{0} = \int_{\RR^{3}} \bigl(-\frac{\delta}{\delta A_{k}(\vec{x})}
	\frac{\delta}{\delta A_{j}(\vec{x})}
	+ \Delta A_{j}(\vec{x}) A_{j}(\vec{x}) \bigr)d^{3}x, 
    \quad \partial_{k} A_{k} = 0 ,
\end{equation*}
    which appears, for example, in electrodynamics or as a result of
    renormalization of a gauge theory, is a limiting case of self-adjoint
    extension of some symmetric operator defined on a restricted set
    of states.
    At the same time generic self-adjoint extensions 
    turn out to be dependent on an extension parameter and for that
    reason do not possess scale invariance.

    Due to the lack of adequate definition of a scalar product on the space
    of functionals, which describe states of the stationary picture of the
    quantum field theory, we shall not make strict statements about the
    self-adjointness or symmetricity.
    Instead, we shall provide a sketch for the new vacuum state and its
    excitations (the Fock space).
    These states satisfy the equations for ``eigenstate'' functionals and
	form a hierarchy of creation and annihilation of particles.
	It is natural to demand that these equations match
	the functional equations
\begin{equation*}
    \HH_{0} \Phi_{\sigma_{n}}(A) = \Lambda_{\sigma_{n}} \Phi_{\sigma_{n}}(A) ,
\end{equation*}
	for eigenstates of Hamiltonian
$ \HH_{0} $,
	but at the same time they could be defined on a set of functionals
	which satisfy different conditions in the vicinity of the ``boundary''
    functions.
	For the role of such boundary points in the configuration space,
	where the boundary conditions of the new functional space are set,
	one can take the functions with singularities behaving as
\begin{equation}
\label{Asing}
    \vec{A}(x) \sim \frac{\vec{A}_{0}}{|x|}, \quad |x| \to 0 ,
	\quad \vec{x} \in \RR^{3} .
\end{equation}
	The possible self-adjoint extensions of the theory, therefore,
    will depend on a
	certain preferred point in the three-dimensional space,
    which should be associated with a localization in the interaction terms.
    However as
	self-interaction is ``turned on'',
	such extensions of the Hamiltonian and the related states will most
	likely turn out to be unstable.
	They, however, may still contribute to the scattering matrix as
	intermediate states for particles interacting via the transverse field.

    For the sake of brevity we shall use
    the following notations for the scalar and vector products
\begin{equation*}
    \vec{A}\cdot\vec{B} = A_{j}B_{j} ,\quad
	(\vec{A}\times\vec{B})_{n} = \epsilon_{njk} A_{j} B_{k} ,
    \quad j,k,n = 1,2,3,
\end{equation*}
    and we shall always assume summation in the repeated indices.

\section{Finite-dimensional examples with
         singular interactions}
    We start with the finite-dimensional examples
    from quantum mechanics
	with the intent to generalize some of their properties
	to the infinite-dimensional case.
	Let
\begin{equation*}
    \Hh_{\ve} = \Delta + \ve \delta(x) = - 
	\frac{\partial^{2}}{\partial x_{k}^{2}}
	    + \ve \delta(x)
\end{equation*}
	be the Hamiltonian of a particle existing in the two- or three-dimensional space
	and interacting with a
$ \delta $-potential centered at the origin.
	Hamiltonian
$ \Hh_{\ve} $
	does not have a correct definition in terms of a closed operator
	in Hilbert space.
	One can, however, consider the action of
$ \Hh_{\ve} $
	on the set of smooth functions falling off towards the origin along with their derivatives.
	This action corresponds to a symmetric operator
\begin{equation*}
    \Hh: \;\; \Hh f(\vec{x}) = \Delta f(\vec{x}) = - 
	\frac{\partial^{2}}{\partial x_{k}^{2}} f(\vec{x}) ,
\end{equation*}
	which, evidently, does not account for potential
$ \ve \delta(\vec{x}) $.
	In terms of the spherical coordinates in two-dimensional
\begin{equation*}
    \vec{x} = \vec{x}(r,\varphi)
    = \begin{pmatrix} r\cos\varphi\\
        r\sin\varphi
        \end{pmatrix}, \quad
    \begin{array}{l}
	0 \leq r,\\ 0 \leq\varphi < 2\pi
    \end{array}
\end{equation*}
	or three-dimensional space
\begin{equation}
\label{sphchange}
    \vec{x} = \vec{x}(r,\psi,\varphi)
    = \begin{pmatrix} r\cos\psi \cos\varphi\\
        r\cos\psi \sin\varphi\\
        r\sin\psi
        \end{pmatrix}, \quad
    \begin{array}{l}
	0 \leq r, \\
	0 \leq\psi \leq\pi,\\
	0 \leq\varphi < 2\pi
    \end{array}
\end{equation}
	the action of operator
$ \Hh_{0} $
	has the following form.
	If a scalar function
$ f(\vec{x}) $
	is represented in terms of a sum of spherical harmonics
$ e^{il\varphi} $ or
$ \YY_{lm}(\psi,\varphi) $
	with coefficients depending on the radial variable,
\begin{gather*}
    f_{2}(\vec{x}) = f_{2}(\vec{x}(r,\varphi)) = \sum_{0\leq l}
    \frac{1}{\sqrt{r}} u_{l}(r)
        \frac{e^{il\varphi}}{\sqrt{2\pi}} , \\
    f_{3}(\vec{x}) = f_{3}(\vec{x}(r,\psi,\varphi)) = \sum_{0\leq |m| \leq l}
    \frac{1}{r} u_{lm}(r)
        \YY_{lm}(\psi,\varphi) , 
\end{gather*}
	then the corresponding operation
$ \Delta $
	acts as follows
\begin{gather*}
    \Delta f_{2}(\vec{x})
        = \sum_{0\leq l} \frac{1}{\sqrt{r}}T_{l-\frac{1}{2}} u_{l}(r)
	    \frac{e^{il\varphi}}{\sqrt{2\pi}} , \\
    \Delta f_{3}(\vec{x})
        = \sum_{0\leq |m| \leq l} \frac{1}{r}T_{l} u_{lm}(r)
	\YY_{lm}(\psi,\varphi) ,
\end{gather*}
	where
\begin{gather}
\label{Tl}
    T_{l} = -\frac{d^{2}}{dr^{2}} + \frac{l(l+1)}{r^{2}} ,\\
\label{Tl1}
    T_{l}^{-1}(r,s) = \frac{1}{2l+1}\bigl(\frac{s^{l+1}}{r^{l}} \theta(r-s)
	+ \frac{r^{l+1}}{s^{l}}\theta(s-r)\bigr).
\end{gather}
	Note that due to the orthonormality of the sets of spherical
	harmonics, each of the scalar products
\begin{equation*}
    (f,g)_{\RR^{2}} = \int_{\RR^{2}} \ol{f(\vec{x})} g(\vec{x}) \,d^{2}x ,
\quad
    (f,g)_{\RR^{3}} = \int_{\RR^{3}} \ol{f(\vec{x})} g(\vec{x}) \,d^{3}x ,
\end{equation*}
	transfers to the coefficient functions
$ u(r) $
	as a plain scalar product on the half-axis
\begin{equation}
\label{plainprod}
    (u,v) = \int_{0}^{\infty} \ol{u(r)} v(r) \, dr .
\end{equation}
	Operators
$ T_{l-\frac{1}{2}} $ and
$ T_{l} $,
	defined on the set of smooth functions vanishing at the origin
	along with their derivatives,
	are essentially self-adjoint with respect to scalar product
(\ref{plainprod})
	at
$ l \geq 1 $.
	At the same time, operators
$ T_{-\frac{1}{2}} $,
$ T_{0} $
    that act on the latter set are symmetric operators with deficiency indices
$ (1,1) $.
	Their self-adjoint extensions
$ T_{-\frac{1}{2}}^{\kappa} $,
$ T_{0}^{\kappa} $
	have continuous spectrum eigenfunctions that look like
\begin{align*}
    v_{\lambda}(r) &= \sqrt{\lambda r} (\alpha_{v\lambda} J_{0}(\lambda r)
	+ \beta_{v\lambda} Y_{0}(\lambda r)) , 
    \quad T_{-\frac{1}{2}}^{\kappa} v_{\lambda} = \lambda^{2} v_{\lambda}\\
    u_{\lambda}(r) &= \alpha_{u\lambda} \sin \lambda r
	+ \beta_{u\lambda} \cos\lambda r , 
    \quad T_{0}^{\kappa} u_{\lambda} = \lambda^{2} u_{\lambda}\\
    &\alpha_{\{u,v\}\lambda} = \alpha_{\{u,v\}}(\lambda,\kappa), \quad
    \beta_{\{u,v\}\lambda} = \beta_{\{u,v\}}(\lambda,\kappa),
\end{align*}
	along with, possibly, some eigenfunctions of the discrete spectrum.
	The actions of extensions
$ T_{-\frac{1}{2}}^{\kappa} $,
$ T_{0}^{\kappa} $
	match the differential operations
$ T_{-\frac{1}{2}} $ and
$ T_{0} $, correspondingly.

	Returning to Cartesian coordinates, therefore, symmetric operators
$ \Hh $
	can be extended to self-adjoint operators
$ \Hh_{2}^{\kappa} $, 
$ \Hh_{3}^{\kappa} $
	defined on the set of functions satisfying the asymptotic conditions
\begin{equation}
\label{fas2}
    \lim_{r\to 0} \frac{f(\vec{x}(r))}{\ln r} = \kappa \lim_{r\to 0}\bigl(
	f(\vec{x}(r)) -\lim_{r'\to 0} \frac{f(\vec{x}(r'))}{\ln r'} \ln r
    \bigr) ,
\end{equation}
	or
\begin{equation}
\label{fas3}
    \lim_{r\to 0} rf(\vec{x}(r)) = -\kappa \lim_{r\to 0}(
	1 + r \frac{\partial}{\partial r} ) f(\vec{x}(r)) ,
\end{equation}
	at the origin
    (see Eqs. (3.43), (3.44) in
\cite{Jackiw}).
	Action of 
$ \Hh_{2}^{\kappa} $,
$ \Hh_{3}^{\kappa} $,
	still matches the sum of squares of the second derivatives
$ \Delta $ on the corresponding set of functions on
$ \RR^{2} $ or
$ \RR^{3} $.

	Extensions
$ \Hh_{2}^{\kappa} $,
$ \Hh_{3}^{\kappa} $
	depend on parameter
$ \kappa $,
	the dimension of which originates from the presence of dimensionality of operator
$ \Hh $:
$ [\Hh] = [x]^{-2} $.
	From physical perspective one can say that
$ \Hh_{2}^{\kappa} $,
$ \Hh_{3}^{\kappa} $
	appear as a result of renormalization of the respective operators
$ \Hh_{\ve} $
	at
$ \ve \to 0 $.
    Meanwhile the singular functions with asymptotics
(\ref{fas2}),
(\ref{fas3})
	at the origin, emerging in the domains of
$ \Hh_{2}^{\kappa} $,
$ \Hh_{3}^{\kappa} $,
    represent the remnant of the
	renormalized singular interaction
$ \ve \delta(\vec{x}) $.
	In the case of a particle in two-dimensional space one
	has the phenomenon of dimensional transmutation ---
	a dimensionless parameter
$ \ve $
	is replaced with a dimensional parameter
$ \kappa $
	during renormalization
\cite{LFBrazil}, \cite{LFres}.

	As another example one can consider two- and three-dimensional
    operators of the type
\begin{equation}
\label{secondex}
    \Delta + \frac{\ve}{|x|^{2}} =
	-\frac{\partial^{2}}{\partial x_{k}^{2}} + \frac{\ve}{|x|^{2}},
\end{equation}
	with
$ \ve $ being a dimensionless parameter.
	Such operators are closed symmetric operators at finite
$ \ve $
	in some vicinity of zero (in two-dimensional case
$ \ve $ has to be positive).
	When drawing a function satisfying the eigenstate equation
	one observes that the increase of the divergence by
$ |x|^{-2} $
	originating from the action of the potential can cancel the
	divergence from the action of the Laplacian.
	Therefore, operator
(\ref{secondex})
	has an alternative basis of locally square-integrable
	``eigenfunctions'' behaving as
$ |x|^{\eta} $
	near the origin
($ \eta = -\sqrt{\ve} $ in two dimensions and
$ \eta = -\frac{1}{2}(1+\sqrt{1+4\ve}) $ in three dimensions),
	that is, it allows self-adjoint extensions
	(this by no means is a mathematically strict explanation
    of the Frobenius method
\cite{Frobenius}).
	One can show that in the limit
$ \ve \to 0 $
	these extensions continuously turn into the corresponding operators
$ \Hh_{2}^{\kappa} $ and
$ \Hh_{3}^{\kappa} $.

\section{Three-dimensional transverse field theory}
	From the perspective of theory of operators in Hilbert space,
	the example of the last section shows that the restriction of the
	domain of Laplacian
$ \Delta $
	to the set of smooth functions vanishing at the origin
	leads to a symmetric operator and an ambiguity in the definition
	of the Hamiltonian of the system.
    At the same time, the interaction, which disappears during
    renormalization,
    only serves as a catalyst for that ambiguity by selecting a preferred
    point in the space.
    In this work we shall try to generalize the experience of the
    finite-dimensional example to the case of field theory.
    We cannot really speak of self-adjointness
    in the case of an inifinite dimensional configurational space
    as long as we do not have a possibility to define the scalar product
    on a wide enough class of functionals.
    For this reason we shall limit ourselves to the description
    of eigenvectors of such systems after renormalization,
    and shall provide an instructive example of an alternative set
    of vacuum state and its excitations.

	Consider the following Hamiltonian functional of the classical
    mechanics
\begin{multline}
\label{qH3}
    \Hcl_{\ve}
    = \iint_{\RR^{3}} E_{j}(\vec{x}) P_{kj}^{\ve}(\vec{x}',\vec{x})
	P_{kj'}^{\ve}(\vec{x}',\vec{y}) E_{j'}(\vec{y})
	    \, d^{3}x \, d^{3}x' \, d^{3}y \,+\\
	+ \int_{\RR^{3}}\bigl((\pl_{k} A_{j}(\vec{x}))^{2}
	+ \ve (A^{3}(\vec{x})+\ldots) \bigr) d^{3}x .
\end{multline}
    where
$ A_{k}^{a}(x) $,
$ E_{k}^{a}(x) $ are the fields of generalized coordinates and their
    conjugate momenta in the three-dimensional space, which satisfy
    the transversality conditions
\begin{equation}
\label{transAE}
    \pl_{k} A_{k}^{a} = 0, \quad 
    \pl_{k} E_{k}^{a} = 0 .
\end{equation}
	We have denoted as 
$ \ve (A^{3}+\ldots) $
	the homogeneous terms of dimension
$ [x]^{-4} $
	of higher order in coordinates
$ A_{k}^{a} $,
    which also include the interaction.
    Matrix
$ P_{kj}^{\ve} $ is the projector from the transverse
	to the covariant-transverse field sets,
\begin{equation*}
    P_{kj}^{\ve}
	= \delta_{kj} - \pl_{k} M^{-1} (\pl_{j}-\ve A_{j}),
	\quad M = (\pl_{j} - \ve A_{j}) \pl_{j} ,
\end{equation*}
    and 
$ \ve $ is a small dimensionless parameter of the theory.
	Fields
$ A_{k}^{a}(x) $,
$ E_{k}^{a}(x) $
	can also have an internal symmetry index
$ a $
	which is everywhere assumed to be summed upon.
	The action of the covariant derivative (and of all objects that contain it)
	may be non-trivial in this index
\begin{equation*}
    (\pl_{k}-\ve A_{k})^{ab} B^{b} = \pl_{k} B^{a}
	- \ve A_{k}^{c} t^{abc} B^{b} .
\end{equation*}
	We shall be only considering the quadratic terms, for which the
    non-triviality in this index reduces to mere summation,
	provided that the matrices 
$ t^{abc} $
	are orthogonal for different
$ c $.
    This way the components corresponding to different values of the upper
    index of field
$ A_{k}^{a}(x) $
    will separate.

	An actual physical example of the Hamiltonian of type
(\ref{qH3})
	is given in the third chapter of book
\cite{FS}.
	Indeed, in Eq.
(2.5)
	therein the following Hamiltonian density is presented,
\begin{equation}
\label{hFS}
    h = \frac{1}{2} (E_{k}^{a})^{2} + \frac{1}{4}
	(\partial_{k}A_{j}^{a} - \partial_{j}A_{k}^{a}
	    - \ve [A_{j},A_{k}]^{a})^{2} ,
\end{equation}
	where
$ \vec{A}(\vec{x}) $ is the transverse field, and
	the conjugate momentum
$ E_{k}^{a} $
	is placed a constraint (2.41) upon,
\begin{equation}
\label{conn}
    (\partial_{k} - \ve A_{k})^{ab} E_{k}^{b} = 0 .
\end{equation}
	After splitting momentum
$ \vec{E}(\vec{x}) $
	into its longitudinal and transverse components
\begin{equation*}
    E_{k} = E_{k}^{L} + E_{k}^{T} ,\quad \partial_{k} E_{k}^{T} = 0,
	\quad E_{k}^{L} = \partial_{k} \xi(\vec{x})
\end{equation*}
	we obtain from condition
(\ref{conn})
	that
\begin{gather*}
    \xi(\vec{x}) = -M^{-1} (\partial_{l} -\ve A_{l}) E_{l}^{T},
    \quad M = (\partial_{j} - \ve A_{j})\partial_{j} , \\
    E_{k} = \bigl(\delta_{kl}
	- \partial_{k} M^{-1} (\partial_{l} -\ve A_{l})\bigr) E_{l}^{T},
\end{gather*}
	and the Hamiltonian density
(\ref{hFS})
	after integrating by parts is transformed to the following
    form of 
Eq.~(\ref{qH3})
\begin{align*}
    h =& \,\frac{1}{2} \bigl( (\delta_{kl}
	- \partial_{k} M^{-1} (\partial_{l} -\ve A_{l}) ) E_{l}^{T} \bigr)^{2}
	+ \frac{1}{2} \bigl( \partial_{k}A_{j}^{a} \bigr)^{2} +\\
	&+ \ve \partial_{k}A_{j}^{a} [A_{j}, A_{k}]^{a}
	+\frac{1}{2} \ve^{2} ([A_{j},A_{k}]^{a})^{2} .
\end{align*}

    The quantum counterpart
$ \HH_{\ve} $ of Hamiltonian
$ \Hcl_{\ve} $
    in the coordinate representation
    acts on functionals
$ \Phi(A_{j}(\vec{x})) $
    in accord with expression
(\ref{qH3})
    with
$ A_{j}(\vec{x}) $
    changed to the operator of multiplication by
$ A_{j}(\vec{x}) $
    and
$ E_{j}(\vec{x}) $
    changed to variation
$ \frac{\delta}{i\delta A_{j}(\vec{x})} $.
    (We assume that the Planck constant equals one and do not discuss
    the ordering of the canonical pairs.)
\begin{multline*}
    \HH_{\ve}
    = - \iint_{\RR^{3}}	P_{kj}^{\ve}(\vec{x}',\vec{x})
	P_{kj'}^{\ve}(\vec{x}',\vec{y})
    \frac{\delta}{\delta A_{j}(\vec{x})} \frac{\delta}{\delta A_{j'}(\vec{y})}
	    \, d^{3}x \, d^{3}x' \, d^{3}y \,+\\
	+ \int_{\RR^{3}}\bigl((\pl_{k} A_{j}(\vec{x}))^{2}
	+ \ve (A^{3}(\vec{x})+\ldots) \bigr) d^{3}x .
\end{multline*}
	During the renormalization procedure when
$ \ve \to 0 $
	the higher order terms
$ \ve (A^{3}+\ldots) $
	disappear, while projector
$ P_{kj}^{\ve} $
	turns into the orthogonal projector onto the transverse component
\begin{equation}
\label{Plim}
    P_{kj}^{\ve} \stackrel{\ve\to 0}{\rightarrow}
	P_{kj} = \delta_{kj} - \pl_{k} \pl^{-2} \pl_{j} ,\quad
    P_{kj}^{T} = P_{kj} ,\quad P_{kn} P_{nj} = P_{kj} .
\end{equation}
	However, in general there is a difference between the result
$ \HH_{\text{ren}} $
    of renormalization of
$ \HH_{\ve} $ at
$ \ve\to 0 $
    and the first term
$ \HH_{0} $ in the expansion of
$ \HH_{\ve} $
    in
$ \ve $
    in the vicinity of zero
\begin{equation}
\label{Hexp}
    \HH_{\ve} = \HH_{0} + \sum_{n\geq 1}
	\frac{\partial^{n} \HH_{\ve}}{\partial\ve^{n}}\bigr|_{\ve =0}
	\frac{\ve^{n}}{n!} .
\end{equation}
    Treatment of divergences in
$ \HH_{\ve} $
    requires introduction of regulariztion parameters,
    which in turn are related to
$ \ve $ (see the resolvent example in
\cite{LFBrazil} or \cite{LFres}).
    As a result some or all terms in expansion
(\ref{Hexp})
    may become finite even at
$ \ve = 0 $
    and the renormalized Hamiltonian
$ \HH_{\text{ren}} $
    will not be equal
$ \HH_{0} $.
    Below we propose that the contribution of these finite terms is such that
    it only forces
$ \HH_{\text{ren}} $
    to acquire an alternative set of eigenstates, while preserving the
    action of
$ \HH_{0} $.
    We have already seen how this picture is realized in the finite-dimensional
    examples in the last section.

    More specifically, the present proposal relates to the fact that    
    Hamiltonian
$ \HH_{\ve} $
	might have singularities via projector
$ P_{kj}^{\ve} $
    around the boundary functions,
	which locally behave as
$ |x|^{-1} $.
	Higher order homogeneous and interaction terms have singularities
    of the same kind.
	In analogy to example
(\ref{secondex}),
	these two types of singularities can cancel each other,
	and in this way supply the renormalized quantum Hamiltonian
	with a domain having new boundary conditions and, correspondingly,
	different spectral properties.

	In order to see this better, let us consider the action
	of the quantum Hamiltonian operator
$ \HH_{0} $ from
(\ref{Hexp})
	upon functionals
$ \Phi(A) $,
\begin{equation}
\label{qH0}
    \HH_{0}\Phi(A) = - \iint_{\RR^{3}} \frac{\delta}{\delta A_{k}(\vec{x})}
	P_{kj}(\vec{x},\vec{y}) \frac{\delta}{\delta A_{j}(\vec{y})}
	    d^{3}x \,d^{3}y \, \Phi(A)
	+ Q(A) \Phi(A) .
\end{equation}
	Here
$ P_{kj} $
	is the projector
(\ref{Plim})
	onto the transverse subspace, and
$ Q(A) $
	is the quadratic form of the Laplace operator
$ \Delta $,
\begin{align}
\label{QA}
    Q(A) =&\, \int_{\RR^{3}} (\pl_{k}A_{j}(\vec{x}))^{2} d^{3}x = \\
\label{QA1}
	&= - \int_{\RR^{3}} A_{j}(\vec{x}) \frac{\pl^{2}}{\pl x_{k}^{2}}
	A_{j}(\vec{x})\, d^{3} x 
	= \int_{\RR^{3}} A_{j}(\vec{x}) \Delta A_{j}(\vec{x}) \, d^{3}x .
\end{align}
    The unnormalized vacuum state of operator
$ \HH_{0} $
    can be constructed as a Gaussian functional
\begin{equation}
\label{Phi0}
    \Phi_{0}(A) = \exp\{-\frac{1}{2}(A,P\Delta^{\frac{1}{2}}PA)\} .
\end{equation}
    And then the
$ n $-particle excitations (the Fock states)
    are
\begin{equation*}
    \Phi_{\sigma_{n}}(A) = \iint
    \sigma_{n}^{j_{1}\ldots j_{n}} (\vec{x}_{1},\ldots \vec{x}_{n})
	b_{j_{1}}(\vec{x}_{1}) \ldots b_{j_{n}}(\vec{x_{n}})
    d^{3}x_{1} \ldots d^{3}x_{n} \Phi_{0}(A) ,
\end{equation*}
	where
$ \sigma_{n} $ are some Bose-Einstein symmetric functions, and
$ b_{j}(\vec{x}) $ are the creation operators from the corresponding pairs
\begin{equation*}
    b_{j} = P_{jk}\bigl(\frac{\delta}{\delta A_{k}}
	- \Delta^{\frac{1}{2}}_{kj'}A_{j'}\bigr) ,\quad
    a_{j} = P_{jk}\bigl(\frac{\delta}{\delta A_{k}}
	+ \Delta^{\frac{1}{2}}_{kj'} A_{j'} \bigr) ,
\end{equation*}
    	of creation and annihilation operators.
	Here quite essential is the fact that projector
$ P $
	commutes with
$ \Delta $,
	and hence, with an arbitrary function thereof ---
	for example with 
$ \Delta^{\frac{1}{2}} $
($ \Delta $ should be defined as a self-adjoint operator).

	Quantum operator
$ \HH_{0} $
	intermixes functionals
$ \Phi_{\sigma_{n}} $,
	however, as is easy to see, 
	it leaves
$ n $-particle subspaces invariant.
	For further diagonalization it is necessary to pass
	to the spectral representation of operator
$ \Delta $,
	which we shall do below in the framework of a more general approach.
	The main idea of that approach is
	to construct an alternative vacuum state via a method
	which, in analogy to the second quantization, can be called
	the method of second self-adjoint extensions.

\subsection{Method of second self-adjoint extensions}
	In the case when the quantum Hamiltonian has the form of 
Eq.~(\ref{qH0}),
	and the positive closed quadratic form
$ Q(A) $
	admits non-trivial extensions,
	there arises a natural way of constructing an alternative
    set of ``eigenstates'' of operator
$ \HH_{0} $.

    In general a closed semi-bounded quadratic form
$ Q(A) $
    can be defined by means of a closed operator
$ S $,
    symmetric or self-adjoint
    in scalar product
$ (\cdot , \cdot ) $,
	via a natural formula
\begin{equation*}
    Q(A) = (A,SA) = (SA,A).
\end{equation*}
	Herein domain
$ \DD_{S} $
	of operator
$ S $
	is contained in domain
$ \DD_{Q} $
	of form
$ Q $,
	and the latter, generally, differs from the former quite significantly.
(One can see, that for the field
$ A $ in
(\ref{QA1})
    to be in the domain of
$ \Delta $
    it should be twice differentiable, while the integral in
(\ref{QA})
    requires the existence only of the first derivative of
$ A $.)
    As long as
$ Q(A) $ 
    is semi-bounded and
$ S $
    is symmetric,
    it allows self-adjoint extensions
$ S_{\kappa} $
	one of which -- Friedrichs extension
\cite{FStone} --
	also defines form
$ Q $,
	while the rest of the extensions, provided they are semi-bounded,
	define different quadratic forms
$ Q_{\kappa} $
(for general material on quadratic forms see section VIII.6 of book
\cite{RS1}).
	These quadratic forms in certain cases (in fact, in a large number
    of simple examples)
	are extensions of the original form
\begin{equation*}
    Q \subset Q_{\kappa} ,
\end{equation*}
	that is, the domain of 
$ Q $
	is contained in the closure of the domain of 
$ Q_{\kappa} $
\begin{equation*}
    \DD_{Q} \subset \ol{\DD}_{Q_{\kappa}} ,
\end{equation*}
	and for all vectors
$ A $ from
$ \DD_{Q} $
	the equality
\begin{equation*}
    Q(A) = Q_{\kappa}(A) ,\quad A\in \DD_{Q} ,
\end{equation*}
	is obeyed.
	In particular, 
\cite{Inv}
	gives spherically symmetric extensions of quadratic form
(\ref{QA})
\begin{equation}
\label{QkA}
        Q_{\kappa}(A) = \lim_{r\to 0}\Bigl(
    \int_{\RR^{3}\setminus B_{r}}
        \bigl|\frac{\partial A_{j}}{\partial x_{k}}\bigr|^{2} d^{3} x -
    (\frac{5}{3r}+ \frac{44}{27}\kappa) \int_{\partial B_{r}}
        |\vec{A}(\vec{x})|^{2} d^{2} s \Bigr) ,
\end{equation}
	for transverse vectors
$ \vec{A}(\vec{x}) $
	with respect to the scalar product
\begin{equation*}
    (\vec{A},\vec{B})_{\RR^{3}} = \int_{\RR^{3}}
	\ol{A_{j}(\vec{x})} B_{j}(\vec{x}) \,d^{3}x .
\end{equation*}
	By
$ B_{r} $ we have denoted a ball of radius
$ r $
	centered at any preferred point.
    For all real-valued vector fields that are regular at that point
    (which we shall further take to be the origin) the value of form
$ Q_{\kappa} $
	obviously equals the value of form
(\ref{QA})
\begin{equation*}
        Q_{\kappa}(A) = Q(A) = \int_{\RR^{3}}
        \bigl(\frac{\partial A_{j}}{\partial x_{k}}\bigr)^{2} d^{3} x .
\end{equation*}
	But the domain of form
$ Q_{\kappa} $
	also includes fields with singularities of type
(\ref{Asing})
	in their three transverse components
	of angular momentum
$ l=1 $.
	The reason for this is that for such fields
	the singularities of the order
$ r^{-1} $
	in the volume integral in
(\ref{QkA})
	get canceled by the singularities of the integral over the boundary of
    ball
$ B_{r} $.
	Notably, the domains of all non-trivial extensions
$ Q_{\kappa} $
	coincide and do not depend on
$ \kappa $.
	We should also add that the coefficient at the dimensional parameter
$ \kappa $
	in equation
(\ref{QkA})
	can be taken arbitrary,
	the value
$ \frac{44}{27} $
	has been chosen to conform to boundary condition
(\ref{cTb}),
	which will be introduced later.

	Next we note that
	by the reason that singular fields of the form
(\ref{Asing})
	are inadmissible for higher order terms of Hamiltonian
$ \HH_{\ve} $
    we can demand that the basic relations for
    ``eigenfunctionals''
$ \Phi_{\sigma_{n}}(A) $
	of operator
$ \HH_{\text{ren}} $
\begin{equation*}
    \HH_{\text{ren}} \Phi_{\sigma_{n}}(A)
	= \Lambda_{\sigma_{n}} \Phi_{\sigma_{n}}(A)
\end{equation*}
	be obeyed only on the domain of quadratic form
$ Q(A) $.
	But on that domain the above relations would also be obeyed
	for a quantum operator with form
$ Q_{\kappa}(A) $
	in place of form
$ Q(A) $.
	Form
$ Q_{\kappa}(A) $,
	after taking the square root of and substituting into a Gaussian integral
	of the type 
(\ref{Phi0}),
	yields a radically different vacuum state and a different set of
	excitations corresponding to a different operator
$ \HH_{\text{ren}}^{\kappa} \neq \HH_{0} $.
	One can propose that operator
$ \HH_{0} $
	is a self-adjoint extension of some symmetric operator
	which is defined on a set of functionals rapidly vanishing near
	boundary functions with singularities of type
(\ref{Asing}).
	This symmetric operator also admits other extensions 
$ \HH_{\text{ren}}^{\kappa} $,
{\it i.e.} the ones
	the ``eigenstates'' for which are built by means of the quadratic form
$ Q_{\kappa}(A) $.

    It should be said that any other (possible) extension of form
(\ref{QA})
    can be used instead of
(\ref{QkA})
    within the present method,
    we have just picked up the one that we know from
\cite{Inv}.
    For a more detailed study let us switch to spherical coordinates
    and single out the subspace of angular momentum
$ l=1 $
    from the field variables.

\subsection{Vector spherical harmonics and separation
	    of variables}
	Using  scalar spherical functions
$ \YY_{lm}(\psi,\varphi) $
	let us introduce the three vector spherical harmonics (VSH)
\cite{VSH}:
\begin{align*}
    \vec{\Upsilon}_{lm}(\Omega) = & \frac{\vec{x}}{r} \YY_{lm} , \quad
        0 \leq l, \quad |m| \leq l, \\
    \vec{\Psi}_{lm}(\Omega) = & \tilde{l}^{-1} r \vec{\pl} \YY_{lm} , \quad
        1 \leq l , \quad |m| \leq l, \quad \tilde{l} = \sqrt{l(l+1)},\\
    \vec{\Phi}_{lm}(\Omega)
	= & \tilde{l}^{-1} (\vec{x} \times \vec{\pl}) \YY_{lm},
        \quad 1 \leq l , \quad |m| \leq l ,
\end{align*}
	which are functions of angular variables
$ \Omega = (\psi,\varphi) $.
	These functions are mutually orthogonal and normalized
    in terms of integration over the sphere,
\begin{align*}
    \int_{\Sph^{2}} \overline{\vec{\Upsilon}_{lm}(\Omega)}
        \vec{\Psi}_{l'm'}(\Omega) d\Omega & = 0 ,\quad
    \int_{\Sph^{2}} \overline{\vec{\Upsilon}_{lm}(\Omega)}
        \vec{\Upsilon}_{l'm'}(\Omega) d\Omega = \delta_{ll'} \delta_{mm'} , \\
    \int_{\Sph^{2}} \overline{\vec{\Upsilon}_{lm}(\Omega)}
        \vec{\Phi}_{l'm'}(\Omega) d\Omega       & = 0 ,\quad
    \int_{\Sph^{2}} \overline{\vec{\Psi}_{lm}(\Omega)}
        \vec{\Psi}_{l'm'}(\Omega) d\Omega = \delta_{ll'} \delta_{mm'} , \\
    \int_{\Sph^{2}} \overline{\vec{\Phi}_{lm}(\Omega)}
        \vec{\Psi}_{l'm'}(\Omega) d\Omega & = 0 ,\quad
    \int_{\Sph^{2}} \overline{\vec{\Phi}_{lm}(\Omega)}
        \vec{\Phi}_{l'm'}(\Omega) d\Omega = \delta_{ll'} \delta_{mm'} .
\end{align*}
	The vector spherical harmonics enable one to uniquely represent
	a vector function
$ \vec{A}(\vec{x}) $
	as three sums
\begin{equation}
\label{fext}
    \vec{A}(\vec{x}) =
        \sum_{0\leq |m| \leq l} y_{lm}(r) \vec{\Upsilon}_{lm} +
        \sum_{l,m} \chi_{lm}(r) \vec{\Psi}_{lm} +
        \sum_{l,m} w_{lm}(r) \vec{\Phi}_{lm} .
\end{equation}
	For brevity we shall from now on assume that summation in indices
$ l,m $
	is always taken in the range
$ 1 \leq l $, 
$ |m| \leq l $
	unless stated otherwise explicitly.
	For each component of expansion
(\ref{fext}),
	when acted upon with Laplacian
$ \Delta $,
	the following separation of variable takes place
\begin{equation*}
    \Delta \bigl(z(r) \vec{Z}_{lm}\bigr) =
-\frac{1}{r^{2}} \frac{\pl}{\plr} r^{2} \frac{\pl}{\plr} z(r) \vec{Z}_{lm}
        + \frac{z(r)}{r^{2}} \Dtph \vec{Z}_{lm}, \quad
            \vec{Z} = \vec{\Upsilon}, \vec{\Psi}, \vec{\Phi} .
\end{equation*}
	The action of the spherical Laplacian
$ \Dtph $
	on the VSH is non-diagonal (for
$ l \geq 1 $)
	but with the given normalization
	it turns out to be symmetric,
\begin{align*}
    \Dtph \vec{\Upsilon}_{lm} &= (2+\tilde{l}^{2}) \vec{\Upsilon}_{lm}
            - 2 \tilde{l} \vec{\Psi}_{lm} ,\\
                  \Dtph \vec{\Psi}_{lm} &= -2 \tilde{l}
\vec{\Upsilon}_{lm}
            + \tilde{l}^{2} \vec{\Psi}_{lm} ,\\
    \Dtph \vec{\Phi}_{lm} &= \tilde{l}^{2} \vec{\Phi}_{lm} .
\end{align*}
	If one imposes the condition of transversality
(\ref{transAE})
	upon a vector function
$ \vec{A}(\vec{x}) $,
	then it will be parametrized by just two sets of functions
$ u_{lm}(r) $,
$ w_{lm}(r) $
	instead of three as in 
(\ref{fext}),
\begin{equation}
\label{Atrexp}
    \vec{A}(\vec{x}) =
        \sum_{l,m} \bigl(\tilde{l}
	    \frac{u_{lm}}{r^{2}} \vec{\Upsilon}_{lm} +
        \frac{u_{lm}'}{r} \vec{\Psi}_{lm} 
    +   \frac{w_{lm}}{r} \vec{\Phi}_{lm} \bigr) .
\end{equation}
    Each of the first two terms in the bracket is not transverse by itself,
    but the terms become so when taken together,
\begin{align}
\label{treq}
    \vec{\pl} &\cdot
\bigl(\tilde{l}\frac{u_{lm}}{r^{2}}\vec{\Upsilon}_{lm}
        +\frac{u'_{lm}}{r}\vec{\Psi}_{lm}\bigr) =\\
\nonumber
    &= \tilde{l} \YY_{lm}
        \bigl( (\frac{u'_{lm}}{r^{2}}-\frac{2u_{lm}}{r^{3}})
        \frac{\vec{x}}{r}\cdot\frac{\vec{x}}{r} 
    + \frac{u_{lm}}{r^{2}} \vec{\pl}\cdot \frac{\vec{x}}{r} \bigr) 
    + \tilde{l}^{-1} u'_{lm} \vec{\pl}\cdot\vec{\pl} \YY_{lm} = 0 .
\end{align}

    The action of the quadratic form of the Laplace operator on a transverse
    field
$ \vec{A}(\vec{x}) $
	written in terms of new variables
$ u_{lm}(r) $,
$ w_{lm}(r) $
	takes the following form (see the corresponding equations in
\cite{Lapl})
\begin{equation*}
        Q(A) = \int_{\RR^{3}} A_{j}(\vec{x}) \Delta A_{j}(\vec{x}) \, d^{3}x 
	= \sum_{l,m}\langle u_{lm},\check{T}_{l}u_{lm}\rangle_{l}
	    + \sum_{l,m}(w_{lm},\check{T}_{l}w_{lm}) ,
\end{equation*}
	where
$ \langle \cdot , \cdot \rangle_{l} $
	is the scalar product inherited from 
$ \RR^{3} $
\begin{equation}
\label{angleprod}
    \langle u, v\rangle_{l} = \int_{0}^{\infty} \bigl(
	\ol{u'(r)}v'(r) + \frac{l(l+1)}{r^{2}} \ol{u(r)}v(r)\bigr) dr ,
    \quad u(0) = v(0) = 0,
\end{equation}
	while the radial part of the Laplace operator
$ \check{T}_{l} $
	and the scalar product
$ (\cdot,\cdot) $
	have been defined in
(\ref{Tl}) and
(\ref{plainprod}).
	For now we are assuming that functions
$ u_{lm}(r) $,
$ w_{lm}(r) $
	are smooth enough and fall off rapidly towards the origin.
	A surprising fact significantly simplifying calculations is that
	product
(\ref{angleprod})
	can be defined as a sesquilinear form of operation
$ T_{l} $
	in scalar product
$ (\cdot,\cdot) $
\begin{equation}
\label{Tprod}
    \langle u,v\rangle_{l} = \int_{0}^{\infty} \ol{u(r)} \bigl(
	-\frac{d^{2}}{dr^{2}}v(r) + \frac{l(l+1)}{r^{2}}v(r) \bigr) dr
	= (u, T_{l}v).
\end{equation}
	In order to avoid confusion between the differential operation
$ T_{l} $
	arising from the scalar product and the radial part of the
    Laplace operator,
	we have denoted the latter as
$ \check{T}_{l} $
	and will keep this notation later.

	The kinetic term of Hamiltonian
(\ref{qH0})
	can be re-written as follows,
\begin{multline}
\label{qHkin}
    - \iint_{\RR^{3}} \frac{\delta}{\delta A_{k}(\vec{x})}
	P_{kj}(\vec{x},\vec{y}) \frac{\delta}{\delta A_{j}(\vec{x})}
	\, d^{3}x\, d^{3}y =\\
    = -\iint\Bigl(
\frac{\delta}{\delta w_{l'm'}(r')} \frac{\delta w_{l'm'}(r')}{
    \delta A_{k}(\vec{x})}
+\frac{\delta}{\delta u_{l'm'}(r')}\frac{\delta u_{l'm'}(r')}{
    \delta A_{k}(\vec{x})} \Bigr) P_{kj}(\vec{x},\vec{y}) \times \\
    \times \Bigl(
\frac{\delta w_{lm}(r)}{\delta A_{j}(\vec{y})}
    \frac{\delta}{\delta w_{lm}(r)}
+\frac{\delta u_{lm}(r)}{\delta A_{j}(\vec{y})} \frac{\delta}{\delta u_{lm}(r)}
    \Bigr) dr\, dr'\, d^{3}x\, d^{3}y .
\end{multline}
	In order for projector
$ P_{kj}(\vec{x},\vec{y}) $
\begin{align*}
    P(\vec{x},\vec{y}) =& \sum_{l,m}
\bigl(\frac{\tilde{l}}{s} \vec{\Upsilon}_{lm}(\Omega)
    - \frac{\pl}{\pl s} \vec{\Psi}_{lm}(\Omega) \bigr) T^{-1}_{l}(s,r)
\bigl(\frac{\tilde{l}}{r} \ol{\vec{\Upsilon}_{lm}(\Omega')}
    + \frac{\pl}{\pl r} \ol{\vec{\Psi}_{lm}(\Omega')} \bigr) +\\
 +& \sum_{l,m} s^{-1} \vec{\Phi}_{lm}(\Omega) \delta(s-r)
	\ol{\vec{\Phi}_{lm}(\Omega')} r^{-1} ,
    \quad \vec{x} = (s,\Omega) , \quad \vec{y} = (r, \Omega')
\end{align*}
	to act as a unit operator on transverse functions,
	let us accept the following parametrization for new variables
$ (u_{lm}, w_{lm}) $
	in terms of 
$ \vec{A} $:
\begin{align}
\label{dphiA}
    w_{lm}(r) &= r\int d\Omega \, \ol{\vec{\Phi}_{lm}(\Omega)}\cdot
	\vec{A}(r,\Omega)
	= \frac{1}{r} \int d^{3}x \, \delta(r-s) \ol{\vec{\Phi}_{lm}(\Omega)} \cdot
	    \vec{A}(\vec{x}) , \\
\nonumber
    u_{lm}(r) &= \int ds \, T_{l}^{-1}(r,s) \int d\Omega \,\bigl(
	\tilde{l}\ol{\vec{\Upsilon}_{lm}(\Omega)}
	    -\frac{\pl}{\pl s}s\ol{\vec{\Psi}_{lm}(\Omega)}
	\bigr) \cdot \vec{A}(s,\Omega) =\\
\label{duA}
    &= \int d^{3}x \,
	\bigl(\frac{\tilde{l}}{s^{2}}T_{l}^{-1}(r,s)
	    \ol{\vec{\Upsilon}_{lm}(\Omega)}
+\frac{1}{s}(\frac{\pl}{\pl s}T_{l}^{-1}(r,s))\ol{\vec{\Psi}_{lm}(\Omega)}
	\bigr) \cdot \vec{A}(\vec{x}) ,
\end{align}
	where again
$ \vec{x} = \vec{x}(s,\Omega) $
    and
$ T_{l}^{-1} $
    come from
(\ref{Tl1}).
	It is not difficult to see that these expressions restore fields
$ u_{lm}(r) $,
$ w_{lm}(r) $ from
$ \vec{A}(\vec{x}) $
	expressed as 
(\ref{Atrexp}),
    while, at the same time, they get annihilated on any longitudinal component
\begin{equation*}
    \vec{A}^{L}(\vec{x}) = \sum_{0\leq |m| \leq l} \bigl(
	v'_{lm}(r)\vec{\Upsilon}_{lm}(\Omega) +\frac{\tilde{l}}{r}v_{lm}(r)
	    \vec{\Psi}_{lm}(\Omega)\bigr)
    = \vec{\partial} \sum_{0\leq |m|\leq l} v_{lm}(r) Y_{lm}(\Omega).
\end{equation*}
	Let us calculate the variations
$ \frac{\delta w_{lm}}{\delta A} $,
$ \frac{\delta u_{lm}}{\delta A} $
	from
(\ref{dphiA}),
(\ref{duA}) and substitute them into
(\ref{qHkin}).
	We find,
\begin{align*}
    -\int& dr\, dr'\, d^{3}x \Bigl(\frac{\delta}{\delta w_{l'm'}(r')}
\vec{\Phi}_{l'm'}(\Omega) \frac{\delta(r'-s)}{r'} \cdot
	\frac{\delta(s-r)}{r} \ol{\vec{\Phi}_{lm}(\Omega)}
	    \frac{\delta}{\delta w_{lm}(r)}
    +\\
&\quad +\frac{\delta}{\delta u_{l'm'}(r')}
    \bigl(\frac{\tilde{l}}{s^{2}} T_{l'}^{-1}(r',s)
	\vec{\Upsilon}_{l'm'}(\Omega)
    +\frac{1}{s}(\frac{\pl}{\pl s}T_{l'}^{-1}(r',s))\vec{\Psi}_{l'm'}(\Omega)
	\bigr)\cdot \\
&\quad\quad \cdot \bigl(\frac{\tilde{l}}{s^{2}}T_{l}^{-1}(s,r)
    \ol{\vec{\Upsilon}_{lm}(\Omega)} +\frac{1}{s}(\frac{\pl}{\pl s}
    T_{l}^{-1}(s,r))\ol{\vec{\Psi}_{lm}(\Omega)} \bigr)
	\frac{\delta}{\delta u_{lm}(r)} \Bigr) =\\
&= -\int dr\, \frac{\delta}{\delta w_{lm}(r)} \frac{\pl}{\pl w_{lm}(r)} -\\
&\quad    -\int dr'\, ds\, dr\, \frac{\delta}{\delta u_{lm}(r')}
    T_{l}^{-1}(r',s)
    (-\frac{\pl^{2}}{\pl s^{2}}+\frac{\tilde{l}^{2}}{s^{2}})
    T_{l}^{-1}(s,r) \frac{\delta}{\delta u_{lm}(r)} ,
\end{align*}
	where we have immediately dropped the cross terms of
$ u $ and
$ w $
	which vanish due to orthogonality of the VSH.
	In the last term the action of
$ T_{l}^{-1} $ on
$ T_{l} $
	produces a
$ \delta $-function
	which removes one integration.
	It is worth noting here that the appearance of coefficient
$ T_{l}^{-1}(r,s) $
	in the square of conjugate ``momenta''
	({\it i.e.} in the kinetic part of the Hamiltonian) is quite natural
	if the ``coordinate'' variable 
$ u_{l}(r) $
	is measured by scalar product
(\ref{Tprod})
	involving operation
$ T_{l} $.

	Adding up the kinetic and potential parts we find the following
	expression for Hamiltonian
(\ref{qH0})
	in terms of the new variables,
\begin{align*}
    \HH_{0} =& \sum_{l,m} \bigl( -\int_{0}^{\infty} dr
	\frac{\delta}{\delta w_{lm}(r)} \frac{\delta}{\delta w_{lm}(r)}
	    + (w_{lm},\check{T}_{l} w_{lm})\bigr)+\\
    &+ \sum_{l,m} \bigl(-\iint_{0}^{\infty} dr dr'
    \frac{\delta}{\delta u_{lm}(r')} T_{l}^{-1}(r',r)
	\frac{\delta}{\delta u_{lm}(r)}
	    + \langle u_{lm}, \check{T}_{l} u_{lm}\rangle_{l} \bigr).
\end{align*}
	As was expected, variables
$ w_{lm} $, 
$ u_{lm} $
	separate for all
$ l $ and $ m $,
	while the vacuum states and excitations of this Hamiltonian
	can be sought as products of states of Hamiltonians
\begin{equation*}
    \HH_{lm} = -\iint_{0}^{\infty} dr dr'
    \frac{\delta}{\delta u_{lm}(r')} T_{l}^{-1}(r',r)
	\frac{\delta}{\delta u_{lm}(r)}
	+ \langle u_{lm}, \check{T}_{l}u_{lm}\rangle_{l}
\end{equation*}
	and
\begin{equation*}
    \HH_{lm}' = -\int_{0}^{\infty} dr
	\frac{\delta}{\delta w_{lm}(r)} \frac{\delta}{\delta w_{lm}(r)}
	    + (w_{lm},\check{T}_{l} w_{lm}) .
\end{equation*}
	Hamiltonians
$ \HH_{lm}' $
	are operators found after switching to spherical coordinates
	and separating the variables in a Hamiltonian for
	a free scalar field.
	For
$ l \geq 1 $
    their eigenvectors are evidently defined unambiguously,
	and so we do not consider them in detail,
	paying close attention to operators
$ \HH_{lm} $ instead.

\subsection{Extensions of quadratic form of operator $ \check{T}_{1} $}
	It was shown in
\cite{Inv}
	that operator
$ \check{T}_{1} $
	in scalar product
$ \langle \cdot , \cdot \rangle_{1} $
	is a symmetric operator with deficiency indices of
$ (1,1) $.
	This operator has non-trivial self-adjoint extensions
	that act as mixed expressions
\begin{equation}
\label{Text}
    \check{T}_{1\kappa} u(r) = T_{1} u(r) - \frac{2}{r} u'(0)
    = -\frac{d^{2}u(r)}{dr^{2}} + \frac{2}{r^{2}} u(r) -\frac{2}{r}u'(0) .
\end{equation}
	on the domains
\begin{equation}
\label{cTb}
    \DD^{\kappa} = \{u(r): \quad \langle u,u\rangle_{1} < \infty, \;
	\langle \check{T}_{1\kappa} u, \check{T}_{1\kappa} u\rangle_{1} <\infty,
	\; 3u''(0) = 4u'(0) \} .
\end{equation}
	Operators
$ \check{T}_{1\kappa} $
have a single-valued continuous spectrum which occupies the non-negative
    half-axis, and to which the following ``eigenfunctions'' (the kernel
    of the spectral transformation) correspond
\begin{equation*}
\label{Tpl}
    p_{1\lambda}^{\kappa}(r)
        = \frac{2r}{\sqrt{2\pi}\lambda^{2}} \frac{d}{dr}\frac{1}{r}
    (\cos(\zeta +\lambda r) - \cos\zeta) ,
\end{equation*}
	where the phase shift
$ \zeta $
    is defined by
\begin{equation*}
    e^{2i\zeta} = \frac{\lambda - i\kappa}{\lambda + i\kappa}.
\end{equation*}
	At
$ \kappa < 0 $
	operator
$ \check{T}_{1\kappa} $
	has an eigenvalue
$ -\kappa $
    of multiplicity one
	(the discrete spectrum) and an eigenfunction
\begin{equation*}
    q(r) = q_{\kappa}(r)
    = \sqrt{-\frac{2}{\kappa^{3}}}
        \bigl(\kappa e^{\kappa r} + \frac{1-e^{\kappa r}}{r}\bigr) .
\end{equation*}
	The set
$ \{p_{1\lambda}^{\kappa}, q \} $
	enjoys the conditions of orthogonality
\begin{equation*}
    \langle p_{1\lambda}^{\kappa} , p_{1\mu}^{\kappa} \rangle_{1}
	= \delta(\lambda-\mu) ,
    \quad \langle p_{1\lambda}^{\kappa} , q \rangle_{1} = 0 ,
    \quad \langle q , q \rangle_{1} = 1
\end{equation*}
	and completeness,
\begin{equation*}
    \int_{0}^{\infty} p_{1\lambda}^{\kappa}(r)
	T_{1}^{s} p_{1\lambda}^{\kappa}(s) \,d\lambda
        + q(r) T_{1}^{s} q(s)\bigr|_{\kappa < 0} = \delta(r-s) ,
\end{equation*}
	where index
$ s $
	of differential operation
$ T_{1}^{s} $
	emphasizes that the latter acts on variable
$ s $.
	Operators
$ \check{T}_{1\kappa} $
	generate extensions
$ \langle u, \check{T}_{1\kappa} u\rangle_{1} $
	of the quadratic forms from the potential parts of Hamiltonians
$ \HH_{1m} $.
	The original form
$ \langle u, \check{T}_{1} u\rangle_{1} $
	is defined on the set of doubly differentiable functions
	vanishing at the origin along with their first derivatives
\begin{equation*}
    \WW^{2}_{0} = \{u(r):\quad \langle u,u\rangle_{1} <\infty, \;
	\langle u, \check{T}_{1} u\rangle_{1} < \infty, \; u(0)=u'(0) =0 \} ,
\end{equation*}
	which corresponds to differentiable fields
\begin{equation}
\label{Atrns}
    \vec{A}(\vec{x}) =
        \sqrt{2}
	    \frac{u_{1m}(r)}{r^{2}} \vec{\Upsilon}_{1m}(\psi,\varphi) +
        \frac{u_{1m}'(r)}{r} \vec{\Psi}_{1m}(\psi,\varphi)
\end{equation}
	regular at the origin.
	The extended forms
$ \langle u, \check{T}_{1\kappa} u\rangle_{1} $
	are defined on a set of functions with an arbitrary bounded value
	of the derivative at the origin
\begin{equation*}
    \WW^{2}_{1} = \{u(r):\quad \langle u,u\rangle_{1} <\infty, \;
	\langle u, \check{T}_{1\kappa}u\rangle_{1} < \infty, \; u(0)=0 \} .
\end{equation*}
	Obviously the latter form equals the former on set
$ \WW^{2}_{0} $
\begin{equation*}
    \langle u, \check{T}_{1\kappa} u\rangle_{1}= \langle u,
	\check{T}_{1} u\rangle_{1} ,\quad u \in \WW^{2}_{0} 
\end{equation*}
    as long as the last term in
(\ref{Text})
    vanishes.
	We shall not provide a symmetric limiting expression like
(\ref{QkA})
    for the extended
	form, as we pass right on to the spectral expansion instead,
\begin{equation*}
    \langle u, \check{T}_{1\kappa} u\rangle_{1}
    = \iint_{0}^{\infty} Q_{\kappa}(r,s) T_{1}^{r} \ol{u(r)}
	T_{1}^{s} u(s) \,dr\,ds ,
\end{equation*}
	with
\begin{equation*}
    Q_{\kappa}(r,s) = \int_{0}^{\infty} p_{1\lambda}^{\kappa}(r)
    	p_{1\lambda}^{\kappa}(s) \lambda^{2} \,d\lambda
        - \kappa^{2} T_{1}^{r} q(r) T_{1}^{s} q(s) \bigr|_{\kappa <0} ,
\end{equation*}
	wherein the second term exists only for
$ \kappa < 0 $.

	To conclude this subsection we note that form
$ \langle u, \check{T}_{1} u\rangle_{1} $
	is in fact a special case of form
$ \langle u, \check{T}_{1\kappa} u\rangle_{1} $
	corresponding to
$ \kappa = \infty $
    ({\it i.e.} the form of Friedrichs or \emph{maximal} extension
    of symmetric operator
$ \check{T}_{1} $).
	From the perspective of the spectral properties of these forms,
	one can observe that the spherical Bessel function
\begin{equation*}
    p_{1\lambda}(r) = \frac{2r}{\sqrt{2\pi}\lambda^{2}}
	\frac{d}{dr}\frac{1}{r} \sin \lambda r ,
\end{equation*}
	appearing in the parametrization of the non-singular transverse field
(\ref{Atrns}),
	is a limiting case of function
$ p_{1\lambda}^{\kappa} $
\begin{equation*}
    p_{1\lambda}(r)  
        = \lim_{\kappa\to\infty}
    \frac{2r}{\sqrt{2\pi}\lambda^{2}} \frac{d}{dr}\frac{1}{r}
    (\cos(\zeta +\lambda r) - \cos\zeta) ,\quad \zeta(\kappa)\to -\frac{\pi}{2}.
\end{equation*}

\subsection{Hamiltonian eigenstates at
\boldmath{$ l=1 $}}
	The spectral expansion that we have obtained in the previous subsection
	now allows us to write the Gaussian functional
$ \phi_{0}^{\kappa}(u) $
	for the extended quantum operator
\begin{equation*}
    \HH_{1m}^{\kappa} = -\iint_{0}^{\infty} dr ds
    \frac{\delta}{\delta u(s)} T_{1}^{-1}(s,r)
	\frac{\delta}{\delta u(r)}
	+ \langle u, \check{T}_{1\kappa}u\rangle_{1}
\end{equation*}
	as the following exponent of an integral operator
\begin{equation*}
    \phi_{0}^{\kappa}(u)
	= \exp\{-\frac{1}{2} \iint Q_{\kappa}^{\frac{1}{2}}(r,s)
	T_{1}^{r} u(r) T_{1}^{s} u(s) \, dr\,ds\} ,
\end{equation*}
	where
\begin{equation*}
    Q_{\kappa}^{\frac{1}{2}}(r,s) = \int p_{1\lambda}^{\kappa}(r)
	p_{1\lambda}^{\kappa}(s)
	\lambda\,d\lambda - i\kappa q(r)q(s) \bigr|_{\kappa <0} .
\end{equation*}
	In this expression we have purposely pulled out differential operations
$ T_{1}^{r} $, 
$ T_{1}^{s} $
	in order to obtain a more smooth kernel
$ Q_{\kappa}^{\frac{1}{2}} $.
	It is not difficult to see that functional
$ \phi_{0}^{\kappa} $
	satisfies the equation
\begin{equation*}
    \HH_{1m}^{\kappa} \phi_{0}^{\kappa}(u)
	= \Lambda_{0}^{\kappa} \phi_{0}^{\kappa}(u) ,\quad
    \Lambda_{0}^{\kappa} = \int_{0}^{\infty} T_{1}^{r}
	Q_{\kappa}^{\frac{1}{2}}(r,r') |_{r=r'} dr
\end{equation*}
	with some infinite eigenvalue
$ \Lambda_{0}^{\kappa} $.
	In order to diagonalize operator
$ \HH_{1m}^{\kappa} $
	let us pass to the spectral representation of the quadratic form,
	{\it i.e.} perform the substitution
\begin{equation*}
    \hat{u}(\lambda) = \int_{0}^{\infty} p_{1\lambda}^{\kappa}(r)
	T_{1} u(r)\,dr, \quad
	\hat{u}_{d} =  \int_{0}^{\infty} q(r) T_{1} u(r) \bigr|_{\kappa<0}
\end{equation*}
	(note that all functions here are real), then
\begin{equation*}
    \HH_{1m}^{\kappa} = \int \bigl(
-\frac{\delta}{\delta \hat{u}(\lambda)} \frac{\delta}{\delta \hat{u}(\lambda)}
	+ \lambda^{2} \hat{u}^{2}(\lambda) \bigr)d\lambda
	- \kappa^{2} \hat{u}_{d}^{2} \bigr|_{\kappa <0} .
\end{equation*}
	This quantum Hamiltonian is associated to the following
	creation and annihilation operators
\begin{equation*}
    \hat{b}(\lambda) = \lambda \hat{u}(\lambda)
	- \frac{\delta}{\delta \hat{u}(\lambda)} ,\quad
    \hat{a}(\lambda) = \lambda \hat{u}(\lambda) 
	+ \frac{\delta}{\delta \hat{u}(\lambda)}
\end{equation*}
	and to a vacuum state
\begin{equation*}
    \hat{\phi}_{0}(\hat{u}) = \phi_{0}(u(\hat{u})) = \exp\{-\frac{1}{2}
	\int_{0}^{\infty} \hat{u}^{2}(\lambda) \lambda \,d\lambda
	+\frac{i\kappa}{2} \hat{u}_{d}^{2} \bigr|_{\kappa < 0}\} .
\end{equation*}
$ n $-particle eigenstates are constructed as integrals with Bose-Einstein
	coefficients
$ \sigma(\lambda_{1},\ldots \sigma_{\lambda_{n}}) $
\begin{equation}
\label{hatphi}
    \hat{\phi}_{\sigma_{n}}(\hat{u}) = \iint
    \sigma_{n} (\lambda_{1},\ldots \lambda_{n}) \,
	\hat{b}(\lambda_{1}) \ldots \hat{b}(\lambda_{n}) \,
    d\lambda_{1} \ldots d\lambda_{n} \hat{\phi}_{0}(\hat{u}) ,
\end{equation}
	and, furthermore, for
$ \kappa < 0 $
	there are states related to the excitations of the discrete spectrum.

\subsection{Eigenstates of the quantum Hamiltonian of a free transverse field}
	The eigenstates of quantum Hamiltonian
$ \HH_{\text{ren}}^{\kappa} $,
    involving the extended quadratic form
(\ref{QkA}),
	are constructed as products of eigenstates of operators
$ \HH_{lm}' $, with
$ 1\leq l, |m| \leq l $,
$ \HH_{lm} $ with
$ 2\leq l, |m| \leq l $ and
$ \HH_{1m}^{\kappa} $.
	For diagonalization of the first two sets of operators one can exploit
	the standard spectral transformation
\begin{equation*}
    \hat{u}_{lm}(\lambda)
	= \int_{0}^{\infty} p_{l\lambda}(r) T_{l} u_{lm}(r)\,dr ,
    \quad \hat{w}_{lm}(\lambda)
	= \int_{0}^{\infty} \lambda p_{l\lambda}(r) w_{lm}(r)\,dr ,
\end{equation*}
	where
$ p_{l\lambda}(r) $
	are some kind of spherical Bessel functions
\begin{equation*}
    p_{l\lambda}(r) = \frac{2r^{l}}{\sqrt{2\pi}\lambda^{l+1}}
	\bigl(\frac{d}{dr}\frac{1}{r}\bigr)^{l} \sin \lambda r .
\end{equation*}
	The corresponding creation and annihilation operators as well as the
	vacuum states look as follows,
\begin{gather*}
    \hat{b}_{lm}(\lambda) = \lambda \hat{u}_{lm}(\lambda)
	- \frac{\delta}{\delta \hat{u}_{lm}(\lambda)} ,\quad
    \hat{a}_{lm}(\lambda) = \lambda \hat{u}_{lm}(\lambda) 
	+ \frac{\delta}{\delta \hat{u}_{lm}(\lambda)} \\
    \hat{b}'_{lm}(\lambda) = \lambda \hat{w}_{lm}(\lambda)
	- \frac{\delta}{\delta \hat{w}_{lm}(\lambda)} ,\quad
    \hat{a}'_{lm}(\lambda) = \lambda \hat{w}_{lm}(\lambda) 
	+ \frac{\delta}{\delta \hat{w}_{lm}(\lambda)} ,\\
    \hat{\phi}_{0}(\hat{u}_{lm}) = \exp\{-\frac{1}{2}
	\int_{0}^{\infty} \hat{u}_{lm}^{2}(\lambda) \lambda \,d\lambda \} ,\\
    \hat{\phi}'_{0}(\hat{w}_{lm}) = \exp\{-\frac{1}{2}
	\int_{0}^{\infty} \hat{w}_{lm}^{2}(\lambda) \lambda \,d\lambda \} .
\end{gather*}
	Diagonalization of Hamiltonian
$ \HH_{1m}^{\kappa} $
	via the transformation
\begin{equation*}
    \hat{u}_{1m}(\lambda) = \int_{0}^{\infty} p_{1\lambda}^{\kappa}(r)
	T_{1} u_{1m}(r)\,dr, 
\end{equation*}
	has been described in the previous subsection.
	It is worth to note here that in a spherically non-symmetric case
	coefficients
$ \kappa $
	may be different for components corresponding to different values
$ m $
	of the third component of the angular momentum.

	In terms of variables
$ \hat{u}_{lm} $,
$ \hat{w}_{lm} $
	we find the resulting Hamiltonian
\begin{equation*}
    \hat{\HH}_{\text{ren}}^{\kappa}
	= \sum_{-1\leq m\leq 1} \hat{\HH}_{1m}^{\kappa}
	+ \sum_{2\leq l, |m|\leq l} \hat{\HH}_{lm}
	+ \sum_{1\leq l, |m|\leq l} \hat{\HH}'_{lm} ,
\end{equation*}
	and the vacuum state
\begin{equation*}
    \Phi_{0}^{\kappa} = \prod_{-1\leq m \leq 1} \phi_{1m}(\hat{u}_{1m}) \times
    \prod_{2\leq l, |m|\leq l} \phi_{lm}(\hat{u}_{lm}) \times
	\prod_{l,m} \phi'_{lm}(\hat{w}_{lm}) ,
\end{equation*}
	while the
$ n $-particle states are obtained from
Eq.~(\ref{hatphi})
	by replacing creation operator
$ \hat{b}(\lambda) $ with an operator
$ c(\lambda) $
    and
$ \hat{\phi}_{0} $
    with
$ \hat{\Phi}_{0}^{\kappa} $
\begin{equation*}
    \hat{\Phi}_{\sigma_{n}}(\{\hat{u}\}) = \iint
    \sigma_{n} (\lambda_{1},\ldots \lambda_{n}) \,
	c(\lambda_{1}) \ldots c(\lambda_{n}) \,
    d\lambda_{1} \ldots d\lambda_{n} \hat{\Phi}_{0}(\{\hat{u}\}) ,
\end{equation*}
    where
$ c(\lambda) $
    can take either of the values
$ \hat{b}_{lm}(\lambda) $ or $ \hat{b}'_{lm}(\lambda) $.

\section{Conclusion and discussion}
	We have constructed a system of states 
	satisfying the eigenstate equations for a quantum Hamiltonian operator
	of a free transverse field.
	The resulting sets generally depend on a preferred point in space and
	do not possess scale invariance ({\it i.e.} depend on a dimensional parameter).
	Our construction has heavily relied upon the properties of extensions
	of the quadratic form of the Laplace operator which appears
    in the potential term of the Hamiltonian.
	These extensions can be written in an invariant form
(\ref{QkA})
	which, similar to the transversality condition, does not imply a transition
	to spherical coordinates or using any preferred functional parametrization
        of the type of
(\ref{Atrexp}).
	A natural question arises here about a possibility to generalize form
(\ref{QkA})
	to the case of two or more preferred points in the space
\begin{equation*}
        Q_{\{\kappa\}}(A) = \lim_{r\to 0}\Bigl(
    \int_{\RR^{3}\setminus \{B_{r,n}\}}
        \bigl(\frac{\partial A_{k}}{\partial x_{j}}\bigr)^{2} d^{3} x -
    \sum_{n=1}^{N}\bigl(\frac{5}{3r}	+ \kappa_{n}\bigr)
	\int_{\partial B_{r,n}} |\vec{A}(\vec{x})|^{2} d^{2} s \Bigr) ,
\end{equation*}
	--- would such a form satisfy the conditions of theorem VIII.15 of
\cite{RS1},
	does it have the corresponding self-adjoint operator, and if it does,
	can a spectral representation be found for the latter?
	A significant difficulty here can be in the transversality condition.
    It is not clear, whether it survives the closure {\it w.~r.} to
    the norm provided by the scalar product and the quadratic form.
	For the case of a single preferred point there already exists
	a spectral representation for the transverse functions,
	which further enables us to discuss related physics,
	while for the case of several preferred points there may be no
	such representation at all.

	Another important remark is that, seemingly, representation
of a physical object (a field mediating an interaction) in terms of a vector function
on a three-dimensional space is not the right method to describe the problem.
	While two functions with singularities at two different points may represent
	a single physical object at different moments in time, they
	cannot be expressed in terms of a common basis,
	{\it i.e.} they do not have a common representation in terms of a single
	orthogonal set.
	This, therefore, creates a significant obstacle for describing the possible
	dynamics of the system.


\section*{Acknowledgments}
    The author is grateful to L.~Faddeev and P.~Bolokhov for discussions.
	The work is supported in part by RFFI grants 14-01-00341, 15-01-03148
	and by ``Mathematical problems of non-linear dynamics'' programme
	of the Russian Academy of Sciences.


\end{document}